**Creative Ideas for Actualizing Student Potential**

Apara Ranjan
Liane Gabora
University of British Columbia

**Abstract**

Human creativity is a multifaceted phenomenon with cognitive, attitudinal, intrapersonal interpersonal, practical, socio-cultural, economic, and environmental aspects (Plucker & Runco, 1999). It can be challenging to incorporate creativity in classrooms (Kampylis, 2008). Teachers tend to inhibit creativity by focusing on correct responses, reproduction of knowledge, and obedience and passivity in class (Alencar, 2002). Right answers, after all, tend to be easier to evaluate than creative ones. Teaching in such a way as to discourage creative answers and approaches may lead to higher scores on standardized tests (with all the sociological, marketplace, and political consequences that entails). Teachers claim to value creativity, but to hold negative attitudes toward, and show little tolerance of, attributes associated with creativity, such as risk taking, impulsivity, and independence (Beghetto, 2006; Fasko, 2001; Runco, 2003; Westby & Dawson, 1995). The majority of teachers express the fear that encouraging creativity in the classroom could lead to chaos (Aljughaiman & Mowrer-Reynolds, 2005; Beghetto, 2007; Westby & Dawson, 1995). When teachers do make efforts to encourage creativity, it is often the case that neither teacher nor students knows what the expectations are. Moreover, students fear that they will be critically judged if they produce something in which they have invested at a personal level.

Despite the potentially threatening aspects of encouraging creative classrooms, we believe that no other investment in education could be more important and potentially rewarding. Both the words student and the word *creativity* often go hand-in-hand with the word *potential*, i.e., *student potential* and *creative potential*. The phrase *creative potential* is generally used to refer to how likely a given individual is to manifest creative works in the future, as assessed by their scores on various creativity tests. The concept of potentiality and its relationship to actuality and context has been studied in depth by physicists. In this chapter we will explore how the physical conception of potentiality can shed light on the notion of potential as it applies in the classroom, and in particular, on how we conceive of and actualize students' creative potential. We then discuss methods that will help optimize the creative potential of students in less threatening ways.

Human creativity is a multifaceted phenomenon with cognitive, attitudinal, intrapersonal interpersonal, practical, socio-cultural, economic, and environmental aspects (Plucker & Runco, 1999). It can be challenging to incorporate creativity in classrooms (Kampylis, 2008). Teachers tend to inhibit creativity by focusing on correct responses, reproduction of knowledge, and obedience and passivity in class (Alencar, 2002). Right answers, after all, tend to be easier to evaluate than creative ones. Teaching in such a way as to discourage creative answers and approaches may lead to higher scores on standardized tests (with all the sociological, marketplace, and political consequences that entails). Teachers claim to value creativity, but to hold negative attitudes toward, and show little tolerance of, attributes associated with creativity, such as risk taking, impulsivity, and independence (Beghetto, 2006; Fasko, 2001; Runco, 2003; Westby & Dawson, 1995). The majority of teachers express the fear that encouraging creativity in the classroom could lead to chaos (Aljughaiman & Mowrer-Reynolds, 2005; Beghetto, 2007; Westby & Dawson, 1995). When teachers do make efforts to encourage creativity, it is often the case that neither teacher nor students knows what the expectations are. Moreover, students fear that they will be critically judged if they produce something in which they have invested at a personal level.

Despite the potentially threatening aspects of encouraging creative classrooms, we believe that no other investment in education could be more important and potentially rewarding. Both the words student and the word *creativity* often go hand-in-hand with the word *potential*, i.e., *student potential* and *creative potential*. The phrase *creative potential* is generally used to refer to how likely a given individual is to manifest creative works in the future, as assessed by their scores on various creativity tests. The concept of potentiality and its relationship to actuality and context has been studied in depth by physicists. In this chapter we will explore how the physical conception of potentiality can shed light on the notion of potential as it applies in the classroom, and in particular, on how we conceive of and actualize students' creative potential. We then discuss methods that will help optimize the creative potential of students in less threatening ways.

## Potentiality

Physicists refer to an entity that is in a full state of potentiality as its *ground state*. The concept is very useful and applicable to entities that could unfold different ways or manifest themselves differently under different contexts or in different environments or situations. In physics, this is the state an entity is in when it is not being measured, because the process of performing the measurement often affects the state of the entity.

The basic notion of ground state has been applied in cognitive psychology, specifically to the study of concepts. Concepts, such as DOG or BEAUTY, are very chameleon-like things, in that they are highly affected by context. Normally, for example, we would think that 'surrounded by water' is a defining feature of the concept ISLAND, but if you use the phrase KITCHEN ISLAND, then hopefully you are not referring to something that is surrounded by water! The concept ISLAND thus can instantly shift to a state in which even a seemingly defining property of islands is not present or expected. The ground state of a concept is the state it is in when you are not thinking about it at all, and thus it is not affected by any context (Gabora & Aerts, 2002; Aerts & Gabora, 2005). A minute ago you were not thinking about a giraffe, and thus your concept GIRAFFE could be said to have been in its ground state. Now you are thinking about giraffes, but it is in the context of reading this book, and so the concept is unavoidably coloured by the context of encountering it while reading this book. When concepts interact with contexts there is always the possibility of new emergent properties that could not be predicted in a straightforward logical way from knowledge of the concept or the context.

We believe that this notion of potentiality applies to the student. Each student can be viewed as a wellspring of potentiality, and the teacher's responsibility is to help this potentiality to actualize or take shape in the context of the classroom environment. This potentiality-driven *perspective* on teaching leads to a different *way* of teaching. When teachers interact with students there is always the possibility of new emergent outcomes that could not be predicted in a straightforward logical way from knowledge of the teacher, the student, or the lesson plan. The thoughts and ideas of the student, as he or she interacts with a particular lesson and approach to teaching it, can potentially follow new, creative trajectories of which the teacher cannot conceive.

## Issues

We begin by discussing some basic issues surrounding incorporating creativity in the classroom. *In order to optimize students' potential in classrooms we need to realize what factors might hinder in actualizing the creative potential of the student in a classroom setting.*

**Expectations**

The term 'creativity' is often used without explanation or establishing clear expectations (Kampylis, 2008). When ideas about creativity are unstated, a teacher's ability to actualize student potential, both positively and negatively, can be affected (Beghetto, 2006; Runco, Johnson, & Bear, 1993). Teachers may benefit from input from psychological research on creativity as they seek to encourage creativity in their classes and establish clear expectations while doing so. Outputs are generally deemed creative if they are original, meaningful, useful or task appropriate, and potentially, possess aesthetic qualities as well (Barron, 1955; Sternberg, Kaufman, & Pretz, 2002). Specifying what kinds of creative outcomes are hoped for, or possible, can help teachers avoid dead-ends, and bewilderment on the part of students. Moreover, creative processes yield not only tangible external outputs, but internal transformative outcomes, including not just enhanced understanding of the creative domain, but enhanced understanding and acceptance more generally, confidence, sense of identity, and personal meaning.

**Flexibility**

One simple way of encouraging creativity in the classroom is to conceive of the students not as receptacles for knowledge but as entities that are biologically predisposed to (amongst other things) to assimilate knowledge, frame it in their own terms, and later, do something with it. As humans we not only want to acquire facts; they want facts that fill in gaps, that prompt a cascade of new understandings that result in more coherent or insightful web of understandings. This can turn learning into a fun and memorable experience. Teachers can help foster this by flexibly adapting their lesson plans to the interests and perspectives of the particular individuals in the classroom, and to spontaneous events that take place in the classroom, or current affairs in the news.

Another simple way of encouraging creativity in the classroom is to give students choice. For example, instead of providing an essay topic, one can invite students to choose a topic related to a general subject. Or, one can offer a selection of possible tasks, to be completed in any order, with the option of focusing exclusively on the task that captures greatest interest. Approaches such as these invite opportunities for students to discover where their unique interests and talents lie. For example, for an assignment in a Psychology of Creativity class, students were given a choice between working on a project or an essay. The instructions were as follows:

> **Details Concerning Presentation, Essay, or Project**
>
> Choose and prepare *either* (1) a presentation, (2) an essay, or (3) a project. To help you choose and research a topic, check out 'Creativity Resources on the Web' and 'Downloadable Papers on Creativity' under 'Student Resources' on Vista. Please type assignments double-spaced, 12-point font, and use APA (American Psychological Association) format for references (examples available under 'Student Resources' on WEB-CT). Assignments MUST be grammatically correct, and will be assessed for content, accuracy, clarity, originality, and strength of arguments. *Essays and projects are due last class before start of class*.

Presentations are done individually or with a partner. Possible topics are listed on WEB-CT or you may choose a chapter of the textbook that we are not covering in class, or your own topic so long as it is relevant to creativity. *All students in the group should arrange to discuss plans with me two weeks prior to the week of their presentation,* and *must be approved by me the week prior*. Creative presentations are encouraged. Have fun with it, and make it fun for the class! (If you do a presentation, you are not obliged **to** turn anything written except the one sentence summary and the one page outline, but if it is in powerpoint please email me the .ppt file.)
An essay can be (1) a critical evaluation of at least two papers on a topic related to creativity that discusses the merits of different techniques, perspectives, or approaches. The papers can be obtained from the resources made available on *Vista*, or they can be articles in a peer-reviewed journal that you find using PSYC-info or Google Scholar. *Or* the essay can be (2) an explanation of how something learned in class applies to or sheds light on your own creative activities. It should be approximately 2000 words (approximately eight pages) not including references.
A project can be anything you want so long as it shows in a nontrivial way that you learned something about creativity in this class. Important: it is not adequate that the project simply *is* creative (e.g. a set of paintings or poems or a scrapbook). It has to demonstrate principles of the psychology of creativity. If you have any uncertainties, come talk to the TA or the Professor early in the semester about possible paper and project ideas.

This assignment is always given on the first day of class, and the final essay or project is due on the last day of class. More detailed instructions are provided in the second class of the semester. These more detailed instructions are provided in Appendix A. During that second class, a list of possible essay and project topics is provided, alongside a list of websites devoted to a multitude of creativity-related topics. Some students who have taken the course in previous years have left behind rather impressive projects that they allow us to show to students as examples of how a project can be tackled.

Assignments such as these are open-ended, and some students may feel lost at first. However, if a student has difficulty coming up with an essay or project topic, he or she can be encouraged to discuss it with the professor during office hours. In our experience all students eventually find a topic that engages them. Another way to help students respond to the open-ended nature of the assignment is to distribute it in sections. For example, the assignment we used in our class was staggered into three phases.

*Assignments with this kind of inherent ambiguity encourage students to relate to distant concepts and look at things from new perspectives*. Studies on conceptual combinations show that dissimilar concepts lead to more creative interpretations (Wilkenfeld & Ward, 2001), suggesting that ambiguous tasks, which might yield more opportunities to connect dissimilar concepts and might produce more creative results. Such an approach encourages bisociation, in which previously unrelated levels of experience or frames of references are suddenly connected (Kostler, 1964). Thus, this approach plays a

major role in creative thinking. Similarly, when students are allowed to choose a topic, they may forge connections between what they are learning and activities that they find personally meaningful.

**Time-Related Factors**

Creating an environment that is conducive to playing, experimenting, trying new things, and thinking things out creatively from new and unusual perspectives, is not readily compatible with a highly structured schedule and strict sets of expectations and goals. Thinking out a creative idea can take time. There are strong neurobiological (Gabora, 2010), experimental (Gabora, 2010; Gabora, O'Connor & Ranjan, in press; Gabora & Saab, 2011), and theoretical (Gabora, 2005) reasons to believe that the generation stage of creative thinking may be divergent not in that it moves in multiple directions or generates multiple possibilities, but in the sense that it produces a raw idea that is vague or unfocused, and that requires further processing to become viable. Similarly, the exploration stage of creative thinking may be convergent, not in the ordinary sense in which an idea is selected from amongst alternatives, but in the sense that a vague idea is considered from different perspectives until it comes into newly defined focus. In other words, the terms *divergent* and *convergent* may be applicable to creative thought in the sense of going from well-defined to ill-defined, and back again. Moreover, it is widely believed that creativity involves shifting back and forth along a spectrum between divergent or associative and convergent or analytic modes of thought, depending on the nature of the problem or task, and the stage one is at of solving or completing it (evidence reviewed in Gabora, 2010; see also Gabora & Ranjan, in press).

A half-baked idea might result when two or more items encoded in overlapping distributions of neural cell assemblies interfere with each other and get evoked simultaneously. This interference has been referred to as *creative interference* because it can lead to creative ideation (Gabora, 2010). When an idea emerges through creative interference, the contributing items are not searched or selected amongst each other because together they form a single structure. This structure can be said to be in a state of potentiality because its ill-defined elements could take on different values depending on how the idea unfolds. This unfolding involves disentangling the relevant features from the irrelevant features by observing how the idea looks from sequentially considered perspectives. In other words, one observes how it interacts with various contexts, either internally generated (think it through) or externally generated (try it out).

The implications of this from an educational standpoint is that to allow 'half-baked' ideas to actualize in to a final creative product that they can feel happy with and proud of, students need sufficient time (Goleman, Kaufman & Ray, 1992). People do not perform well on creative thinking tasks in time-pressured situations. An added benefit of providing sufficient time for creative thinking is that by encouraging students to put more of themselves into the projects they undertake, they often become more intrinsically motivated to complete them (Deci & Ryan, 1985; Deci, 1975; Deci, 1981; Amabile, 1983, 1988, 1990).

A practical way for providing students with opportunities for their ideas to mature and their creative potential to be realized is to give them projects that involve multiple steps or phases, and stagger feedback across these multiple phases. This lets students incubate an idea, or 'sleep on it.' For example, one effective means we have found for doing this involves breaking the project or essay assignment mentioned above into three phases. The instructions for completing the first phase are as follows:

ASSIGNMENT 1 (due Sept. 27, 2011)

State whether you will be doing an essay, project, or presentation. Give the title of it, one sentence describing it, and one scholarly journal reference that you will use to research it. It should be grammatically correct and free of spelling errors. You must include at a minimum one reference to a recent (year 2000 or later) scholarly journal article in APA style.

Submit it through dropbox (do not include an attachment).

EXAMPLE OF HOW TO DO ASSIGNMENT 1:

PSYO 317
Assignment 1
Due: September 10, 2011
Jane Sunbeam
Student Number 1111111

ESSAY

Title:
Neanderthal Creativity

Sentence:
This essay will discuss evidence for artistic playfulness and artistic artifacts that have been found in Neanderthal settlements, and it will address the question of to whether creative abilities and artistic once thought to be uniquely human can be attributed to the Neanderthals.

Reference:
Fitch, W. T. (2006). The biology and evolution of music: A comparative perspective. *Cognition, 100,* 173-215.

Creativity thrives in situations that involve a combination of freedom and constraint. We find that, for relatively unstructured tasks such as this one, providing one or more examples is a useful way to ensure that students know what is expected and have a sense of how to turn abstract instructions into something concrete.

The instructions for completing the second phase are as follows:

ASSIGNMENT 2
One Page Summary of Presentation, Essay, or Project
Due Nov. 10, 2011

The goal of this assignment is to show that you are making progress with your essay or project, and to show that you are able to coherently explain how it relates to the psychology of creativity.

This assignment should include *everything assignment 1 included* but instead of one sentence about your essay or project it should include a <u>full page</u> of <u>double spaced</u> text in <u>full sentences</u> (not point form). The text should introduce the topic and summarize what approach you will take or what you plan to do in your project or essay. It should be grammatically correct and free of spelling errors. You must include at a minimum the one reference you had for your first assignment; it does not hurt to have added a few more references. As with Assignment 1, the reference(s) must be in APA style.

The third phase of the assignment is to write the final essay or create the final project. Note that throughout the process students are encouraged to sense the potential of what they have done to develop into something more. The sense that it holds potential, as well as its capacity to actualize in different ways depending on how one thinks it through, was communicated to the students by giving feedback at each phase of the assignment.

**Creative Teaching Methods**

We now discuss a few of the methods that are useful for teaching creatively and helping to actualize the creative potential of the students.

**Humor**

Humor helps to create a fun and interesting learning environment which not only helps in gaining the attention of the students but also fosters creative thinking. Humor helps break down communication and learning barriers between teachers and students (Berk, 1998). Students tend to rehearse and remember class material more if it is presented with humor. This humor helps to retain their attention. Humor also helps to increase comprehension and cognitive retention mainly due to its reduction of stress and anxiety (Hill, 1988).

Classroom humor can obviously take the form of jokes, but there are other subtle and effective ways of incorporating humor in classroom. It can take the form of quotations or cartoons in lectures, or humorous definitions. For example, the following was used at the end of the class on psychological disorders:

> If you get the joke below you've understood this material…
>
> Welcome to the Psychological Clinic! If you have an obsessive-compulsive disorder, please press 1, as many times as you like. If you are co-dependent, please ask someone to press 2 for you. If you have multiple personality disorder, please press 3, 4, 5, and 6. If you are paranoid or delusional, we know who you are and what you want. Just stay on the line so we can trace the call. If you are schizophrenic, listen carefully and a little voice will tell you which number to press. If you are depressed, it doesn't really matter which number you press. Probably no one will answer anyway.

Humorous material used at the beginning of the class encourages students to come on time and creates a positive, receptive atmosphere that is conducive to learning (Berk, 1998).

**Interactive Classrooms**

The second technique is to incorporate activities that bring the lesson to life. Incorporating group discussions and group activities are commonly used methods of increasing class participation. Note however that group activities are not always conducive to creative thinking, and that people are not necessarily aware of this, as they tend to over-rate ideas produced by a group (Baumeister & Bushman, 2010). Groupwork followed by solitary brainstorming and communicating via networked computers can reduce the chance of blocking anyone from speaking, while making it easier to ignore irrelevant input from others when one is "on a roll" on one's own**.** *We have realized that it is beneficial to make students work on a project intially by themselves and then put the students in groups at a later stage in the project. This provides them a chance to originally come up with an idea and then hone the idea to actualize its potential and transform it into a final product with the help of group discussions and feedback given by others.*

**Personalizing the Learning Experience**

Another technique is to relate material to the students' experience. Encouraging students to share their personal experiences in regard to the topic being discussed helps them personalize the discussion. Also, we can design assignments in such a way that they give students a chance to relate with them in a personal way. This way of designing of assignments may enhance students' intrinsic motivation to learn. For example, after a lesson on developmental life stages, we gave a bonus assignment to the students along with other regular assignments.We asked them to work on a lifespan scrapbook which described the stages of development with examples. This scrapbook either had to describe their lifespan, or the lifespan of someone they knew.

**Metaphor**

The fourth technique is the use of metaphors in teaching. Using metaphors helps in building relations between what the students already know and the new concept being learnt (Glynn & Takahashi, 1998). This method also makes it easier for the students to remember the complex concepts. The enacting of metaphors can prime the existing knowledge (Williams & Bargh, 2008). It has been shown that physical movements (backward, forward) appear to cue memories for past events or thoughts about future events (Miles, Nind, & Macrae, 2010). This research indicates that appropriate use of metaphors and their enactment in the class can help students to make stronger associations with the learned material. The use of animations to explain ideas and making students act out certain ideas helps them to remember them better. For example, while teaching a class on the various different "talk-therapy" approaches to treating mental disorders, we have tried inviting the students to find a partner and have one person act as the therapist and the other act as the client, and try out each of the therapies. This approach has proven useful for helping students remember the material better and become personally engaged in the topic. Some advocates of performance based teaching suggest that the teachers are the stage directors, with the students as the actors (Smith, 1979).

## Conclusion

Creativity has mostly been looked at as a means to an end in education (*i.e.*, improving teaching skills, problem solving, and enhancing motivation). Education systems fear embracing creativity as it threatens the long established relationship between teacher and student and the way classrooms function. There are several challenges in bridging the gap between those involved in education and the creativity researchers. This chapter focused on ways to bridge the gap between creativity research and education methods and proposed ways to optimize student's creative potential. The physical conception of potentiality can shed light on the notion of potential as it applies in the classroom. It is essential to focus on the creative process rather than just on the creative product when we want to foster creativity in students.

Westby, E. L., & Dawson, V. L. (1995). Creativity: Asset of burden in the classroom? *Creativity Research Journal, 8*(1), 1–11.

Williams, L. E., & Bargh, J. A. (2008). Experiencing physical warmth promotes interpersonal warmth. *Science, 322*, 606-607.
14

APPENDIX A

PSYO 317 'Psychology of Creativity'

REQUIREMENTS FOR ALL ESSAYS AND PROJECTS

*Note: It is important that you talk to me in person about your topic well before the submission date if there is anything that is not clear.*

STYLE AND FORMAT

- Must be submitted *on the due date*: *in hardcopy* and also in electronic format by vista dropbox or by vista email (I will let you know)
- Do not call your electronic file 'Creativity.doc'. Electronic form must be labeled in the following format: Last name _ first name _ PSYO317PPE
- *Title page* with title and your name and name of class
- All written documents should be double-spaced; do not justify right-hand margins
- Suggested length: 2000 words (= about 10 pages including cover page and references page) or the equivalent
- Must be free of spelling and grammatical errors
- Should not be worded in an awkward manner; read it over out loud to yourself before submitting to make sure
- *References* listed at end in APA format; there must be at least a few *scholarly* references (e.g. academic journal articles)

CONTENT

- Regarding your choice of topic: it is best to choose something you are passionate about. But if you choose a topic that most people would consider to be only minimally creative (such as, say, hockey) you may have a challenge ahead of you explaining why you think it's creative. If you genuinely believe that most people just haven't yet appreciated the creative element of your chosen topic, and you think you can argue this convincingly, then go for it. But if the reality is that you're just trying to find a way of getting marks for a course while writing about your favourite topic, that will show through.
- Must contain *original work*. (Be aware that there are computer programs that professors can run an essay through to see if it matches anything on the internet.)
- Links to YouTube videos are not original work and will not count as part of your submission, but you can include them if they are useful to illustrate a point.
- Must have *references to scholarly material* including not just books but journal articles. (References to non-academic items are not sufficient although you may add some if they are relevant.)
- Cite *direct source* of literature; do not cite someone who discussed the study but the person who did the study
- If you stick to referencing papers we covered in class you will not get a top mark; additional research is required using PsycInfo or Google Scholar

ESSAYS

- Essays will be assessed for primarily for content, accuracy, clarity, and strength of arguments. Originality, insight, and effort invested are also taken into account.
- Should have introduction that gets reader's attention and states clearly the central idea or focus of the essay
- Body of essay should develop the idea showing clearly a depth of understanding gained by reading articles related in some way to the psychology of creativity
- Should have a concluding paragraph
- Even if your essay is a critical evaluation of a topic discussed in one particular paper, it should be clear that you are using knowledge acquired through reading *other* papers to bear on the evaluation. Similarly, even if you discuss your own creative process, your essay should demonstrate familiarity with the scholarly literature on this creative domain and how it sheds light on your creative process

PROJECTS

- I am extremely open to (even enthusiastic about!) something creative or even outright weird, so long as it isn't just creative but also demonstrates an understanding of the *psychology of creativity*. (e.g. A scrapbook of your trip to Maui will not get a good mark.)
- Integrate the creative output with what you've learned about the creative process through not just the class but also independent scholarly research for this project
- The application of scholarly theories about creativity to particular creative works takes time. Don't worry if you don't see them right away. The best thing you can do is read lots of articles over the course of the semester, and the connections will slowly start to dawn on you. If you put all your effort into the creative project and leave the scholarly part until the last minute, any connections you try to make between creativity theory and the creative work will probably feel forced.
- Must have all parts submitted *at the same time*, and *attached together* or *in a container* so that they cannot be separated.
- If it's a movie, edit out stuff that is not interesting (like asking someone on the street what creativity is and having that person say they don't know), and don't film yourself reading class notes
- Don't make statements such as "this is clearly creative". (Such statements often follow discussion of things that are *not* particularly creative.) Explain why it is creative and what theories it exemplifies or principles it demonstrates in a way that is not gratuitous but genuine.
- **Important:** Projects that consist of fragmentary components can come across as scattered; they *must* at the very least include a synthesis, which explains how they relate to each other, and to the psychology of creativity. A collection of scattered quotes or statements can indicate a fragmented understanding of the topic. Ideally your project should show that you have delved deep into your topic, both in terms of researching it and in terms of synthesizing the research and mulling it over, and arrived at a nuanced understanding of the topic. You should show how research on creativity pertains to you or a particular creative individual or creative task.